\documentclass[epj]{webofc}
\usepackage[varg]{txfonts}   
\usepackage{parskip} 
\begin{document}
\title{The LPSZ-CLASH galaxy cluster sample: combining lensing and hydrostatic mass estimates}
%
%

\author {\firstname{M.}~\lastname{Mu\~noz-Echeverr\'{\i}a}\inst{1}\fnsep\thanks{\email{miren.munoz@lpsc.in2p3.fr}\inst{\ref{LPSC}}}
   \and \firstname{R.}~\lastname{Adam} \inst{\ref{LLR}}
   \and  \firstname{P.}~\lastname{Ade} \inst{\ref{Cardiff}}
    \and  \firstname{H.}~\lastname{Ajeddig} \inst{\ref{CEA}}
   \and  \firstname{P.}~\lastname{Andr\'e} \inst{\ref{CEA}}
   \and \firstname{M.}~\lastname{Arnaud} \inst{\ref{CEA}}
  \and \firstname{E.}~\lastname{Artis} \inst{\ref{LPSC}}
   \and  \firstname{H.}~\lastname{Aussel} \inst{\ref{CEA}}
   \and  \firstname{I.}~\lastname{Bartalucci} \inst{\ref{Milano}}
   \and  \firstname{A.}~\lastname{Beelen} \inst{\ref{IAS}}
   \and  \firstname{A.}~\lastname{Beno\^it} \inst{\ref{Neel}}
   \and  \firstname{S.}~\lastname{Berta} \inst{\ref{IRAMF}}
   \and  \firstname{L.}~\lastname{Bing} \inst{\ref{LAM}}
   \and  \firstname{O.}~\lastname{Bourrion} \inst{\ref{LPSC}}
   \and  \firstname{M.}~\lastname{Calvo} \inst{\ref{Neel}}
   \and  \firstname{A.}~\lastname{Catalano} \inst{\ref{LPSC}}
   \and  \firstname{M.}~\lastname{De~Petris} \inst{\ref{Roma}}
   \and  \firstname{F.-X.}~\lastname{D\'esert} \inst{\ref{IPAG}}
   \and  \firstname{S.}~\lastname{Doyle} \inst{\ref{Cardiff}}
   \and  \firstname{E.~F.~C.}~\lastname{Driessen} \inst{\ref{IRAMF}}
   \and  \firstname{A.}~\lastname{Ferragamo} \inst{\ref{Roma}}
   \and  \firstname{A.}~\lastname{Gomez} \inst{\ref{CAB}}
   \and  \firstname{J.}~\lastname{Goupy} \inst{\ref{Neel}}
   \and  \firstname{F.}~\lastname{K\'eruzor\'e} \inst{\ref{LPSC}}
   \and  \firstname{C.}~\lastname{Kramer} \inst{\ref{IRAME}}
   \and  \firstname{B.}~\lastname{Ladjelate} \inst{\ref{IRAME}}
   \and  \firstname{G.}~\lastname{Lagache} \inst{\ref{LAM}}
   \and  \firstname{S.}~\lastname{Leclercq} \inst{\ref{IRAMF}}
   \and  \firstname{J.-F.}~\lastname{Lestrade} \inst{\ref{LERMA}}
   \and  \firstname{J.-F.}~\lastname{Mac\'ias-P\'erez} \inst{\ref{LPSC}}
   \and  \firstname{A.}~\lastname{Maury} \inst{\ref{CEA}}
   \and  \firstname{P.}~\lastname{Mauskopf}
\inst{\ref{Cardiff},\ref{Arizona}}
   \and \firstname{F.}~\lastname{Mayet} \inst{\ref{LPSC}}
   \and  \firstname{A.}~\lastname{Monfardini} \inst{\ref{Neel}}
   \and  \firstname{A.}~\lastname{Paliwal} \inst{\ref{Roma}}
   \and  \firstname{L.}~\lastname{Perotto} \inst{\ref{LPSC}}
   \and  \firstname{G.}~\lastname{Pisano} \inst{\ref{Cardiff}}
   \and  \firstname{E.}~\lastname{Pointecouteau} \inst{\ref{Toulouse}}
   \and  \firstname{N.}~\lastname{Ponthieu} \inst{\ref{IPAG}}
   \and  \firstname{G.~W.}~\lastname{Pratt} \inst{\ref{CEA}}
   \and  \firstname{V.}~\lastname{Rev\'eret} \inst{\ref{CEA}}
   \and  \firstname{A.~J.}~\lastname{Rigby} \inst{\ref{Cardiff}}
   \and  \firstname{A.}~\lastname{Ritacco} \inst{\ref{ENS},\ref{IAS}}
   \and  \firstname{C.}~\lastname{Romero} \inst{\ref{Pennsylvanie}}
   \and  \firstname{H.}~\lastname{Roussel} \inst{\ref{IAP}}
   \and  \firstname{F.}~\lastname{Ruppin} \inst{\ref{MIT}}
   \and  \firstname{K.}~\lastname{Schuster} \inst{\ref{IRAMF}}
   \and  \firstname{S.}~\lastname{Shu} \inst{\ref{Caltech}}
   \and  \firstname{A.}~\lastname{Sievers} \inst{\ref{IRAME}}
   \and  \firstname{C.}~\lastname{Tucker} \inst{\ref{Cardiff}}
   \and  \firstname{G.}~\lastname{Yepes} \inst{\ref{Madrid}}
}

   \institute{
     Univ. Grenoble Alpes, CNRS, LPSC-IN2P3, 53, avenue
des Martyrs, 38000 Grenoble, France
     \label{LPSC}
     \and
     LLR, CNRS, École Polytechnique, Institut Polytechnique de Paris,
Palaiseau, France
     \label{LLR}
     \and
     School of Physics and Astronomy, Cardiff University, Queen’s
Buildings, The Parade, Cardiff, CF24 3AA, UK
     \label{Cardiff}
     \and
     AIM, CEA, CNRS, Universit\'e Paris-Saclay, Universit\'e Paris
Diderot, Sorbonne Paris Cit\'e, 91191 Gif-sur-Yvette, France
     \label{CEA}
     \and
     INAF, IASF-Milano, Via A. Corti 12, 20133 Milano, Italy
     \label{Milano}
     \and
     Institut d'Astrophysique Spatiale (IAS), CNRS, Universit\'e Paris
Sud, Orsay, France
     \label{IAS}
     \and
     Institut N\'eel, CNRS, Universit\'e Grenoble Alpes, France
     \label{Neel}
     \and
     Institut de RadioAstronomie Millim\'etrique (IRAM), Grenoble, France
     \label{IRAMF}
     \and
     Aix Marseille Univ, CNRS, CNES, LAM, Marseille, France
     \label{LAM}
     \and
     Dipartimento di Fisica, Sapienza Universit\`a di Roma, Piazzale
Aldo Moro 5, I-00185 Roma, Italy
     \label{Roma}
     \and
     Univ. Grenoble Alpes, CNRS, IPAG, 38000 Grenoble, France
     \label{IPAG}
     \and
     Centro de Astrobiolog\'ia (CSIC-INTA), Torrej\'on de Ardoz, 28850
Madrid, Spain
     \label{CAB}
     \and
     Instituto de Radioastronom\'ia Milim\'etrica (IRAM), Granada, Spain
     \label{IRAME}
     \and
     LERMA, Observatoire de Paris, PSL Research University, CNRS,
Sorbonne Universit\'e, UPMC, 75014 Paris, France
     \label{LERMA}
     \and
     Univ. de Toulouse, UPS-OMP, CNRS, IRAP, 31028 Toulouse, France
     \label{Toulouse}
     \and
     Laboratoire de Physique de l’\'Ecole Normale Sup\'erieure, ENS, PSL
Research University, CNRS, Sorbonne Universit\'e, Universit\'e de Paris,
75005 Paris, France
     \label{ENS}
     \and
     Department of Physics and Astronomy, University of Pennsylvania,
209 South 33rd Street, Philadelphia, PA, 19104, USA
     \label{Pennsylvanie}
     \and
     Institut d'Astrophysique de Paris, CNRS (UMR7095), 98 bis boulevard
Arago, 75014 Paris, France
     \label{IAP}
     \and
     Kavli Institute for Astrophysics and Space Research, Massachusetts
Institute of Technology, Cambridge, MA 02139, USA
     \label{MIT}
     \and
     School of Earth and Space Exploration and Department of Physics,
Arizona State University, Tempe, AZ 85287, USA
     \label{Arizona}
     \and
     Caltech, Pasadena, CA 91125, USA
     \label{Caltech}
     \and
     Departamento de F\'isica Te\'orica and CIAFF, Facultad de Ciencias,
Modulo 8, Universidad Aut\'anoma de Madrid, 28049 Madrid, Spain
     \label{Madrid}
   }

\abstract{%
Starting from the clusters included in the NIKA sample and in the NIKA2 Sunyaev-Zel'dovich Large Program (LPSZ) we have selected a sample of six common objects with the Cluster Lensing And Supernova survey with Hubble (CLASH) lensing data. For the LPSZ clusters we have at our disposal both high-angular resolution observations of the thermal SZ with NIKA and NIKA2 and X-ray observations with \textit{XMM-Newton} from which hydrostatic mass estimates can be derived. In addition, the CLASH dataset includes lensing convergence maps that can be converted into lensing estimates of the total mass of the cluster.
One-dimensional mass profiles are used to derive integrated mass estimates accounting for systematic effects (data processing, modeling, etc.).
Two-dimensional analysis of the maps can reveal substructures in the cluster and, therefore, inform us about the dynamical state of each system. Moreover, we are able to study the hydrostatic mass to lensing mass bias, across different morphology and a range of redshift clusters to give more insight on
the hydrostatic mass bias. The analysis presented in this proceeding follows the study discussed in \cite{ferragamo}.
}
\maketitle
\section{Introduction}
\label{intro}
Clusters of galaxies have been shown to be excellent cosmological probes \cite{boquet,planck1,lubini}. Nevertheless, cluster-based cosmological results show a trend towards lower $\sigma_{8}$ than CMB-based results \cite{pratt}. This may be due to a lack of knowledge on cluster physics and its impact on the reliable evaluation of cluster masses. Different mass estimates are affected by different systematic uncertainties \cite{pratt},
so combining observables may help building a consistent and more accurate mass estimate. In this work we will combine hydrostatic equilibrium and lensing mass estimates using the common sample between NIKA, the LPSZ and CLASH.

The NIKA2 Sunyaev-Zel'dovich Large Program (LPSZ) \cite{LPSZ} studies 45 high redshift (z in 0.5-0.9) galaxy clusters selected in SZ. One of its objectives is to reconstruct the hydrostatic equilibrium (HSE) mass of the clusters by combining the electron pressure reconstruction using the thermal SZ (tSZ) effect with the NIKA2 camera \cite{adam1,calvo,NIKA2-electronics,perotto} and the electron density and temperature from X-rays with \textit{XMM-Newton} satellite. The CLASH collaboration \cite{postman} observed 25 massive galaxy clusters of redshift 0.2 to 0.9. From the combined strong and weak lensing analysis, they reconstructed convergence maps ($\kappa$-maps) that are publicly available \cite{zitrin1} and are the starting point for the lensing study in this work.

\section{The LPSZ-CLASH sample}

\begin{figure}
    \centering
    \includegraphics[trim={0cm 0.5cm 0cm 2cm},scale=0.2]{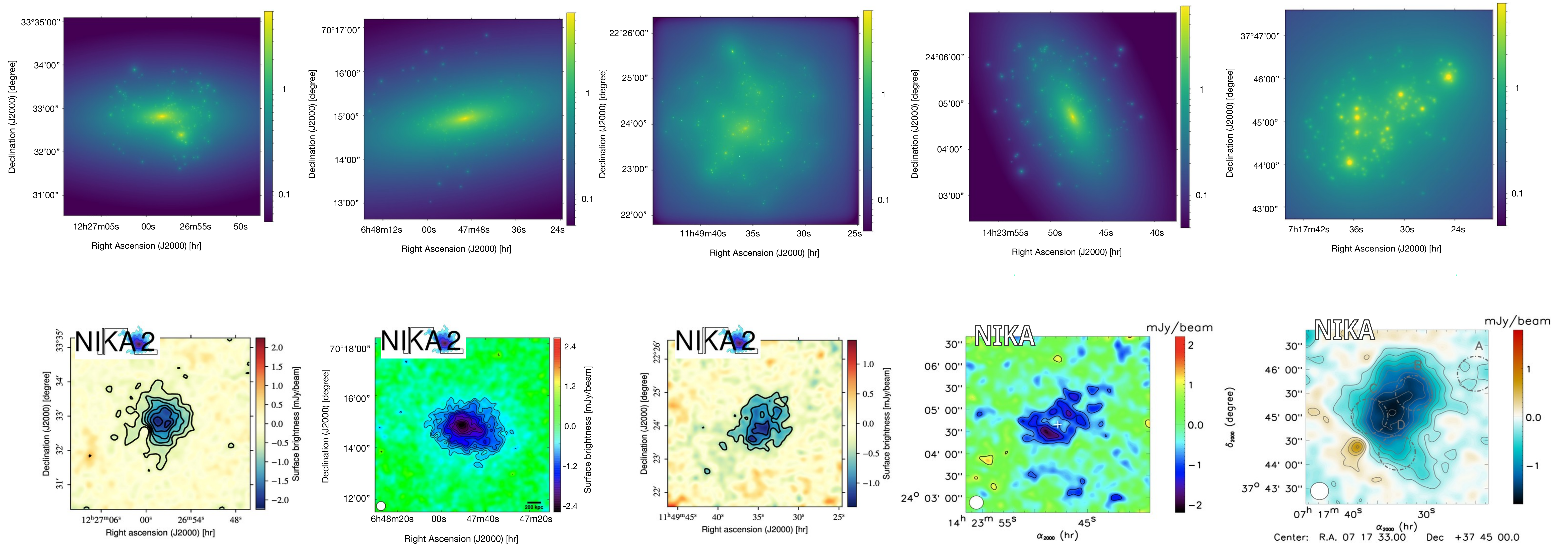}
    \caption{The LPSZ-CLASH sample, from left to right: PSZ2-G160.83+81.66, PSZ2-G144.83+25.11, PSZ2-G228.16+75.20, MACS J1423.8+2404, MACS J0717.5+3745. CLASH $\kappa$-maps on top. NIKA2 and NIKA SZ-maps at the bottom.}
    \label{sample}
\end{figure}

We present in Figure~\ref{sample} five of the six clusters in our sample. The study of the LPSZ-CLASH cluster PSZ2-G045.32-38.46 is ongoing and it will not be presented in this work. PSZ2-G160.83+81.66 (z=0.89), PSZ2-G144.83+25.11 (z=0.58) and PSZ2-G228.16+75.20 (z=0.55) are part of the LPSZ sample and we will study their HSE and lensing mass profiles
. For PSZ2-G144.83+25.11 we will use the results from the thermal SZ and X-ray analysis done in \cite{ruppin1}. Furthermore, MACS J1423.8+2404 (z=0.55) and MACS J0717.5+3745 (z=0.55) were part of the NIKA sample \cite{catalano}. For the HSE mass analysis of MACS J1423.8+2404 we will use the results from \cite{adam2} and the tSZ-map for the study of MACS J0717.5+3745 will be taken from \cite{adam3}. 
\section{Mass profiles reconstruction}
\subsection{Hydrostatic mass profiles}
\label{hse}
The HSE mass reconstruction was done, on the one side, by combining the electron pressure profile obtained from SZ data and the electron density from X-rays as explained in \cite{ruppin1, keruzore, adam2}. On the other side, we combine \textit{XMM-Newton} X-ray spectroscopic temperature and electron density profiles \cite{bartalucci}. For PSZ2-G160.83+81.66 and PSZ2-G228.16+75.20, the contamination of point sources was estimated as in \cite{keruzore} and taken into account for the electron pressure fit on the SZ map. For PSZ2-G144.83+25.11 the southwestern over-pressure region was masked to fit the pressure profile as well as the electron density \cite{ruppin1}. The tSZ signal in MACS J1423.8+2404 is very contaminated by point sources and they were subtracted before the pressure profile fit (M2 method in \cite{adam2}).
\subsection{Lensing mass profiles}
For the lensing mass estimation we followed the procedure described in \cite{ferragamo}. Convergence maps ($\kappa$-maps) are surface mass density maps ($\Sigma$) in critical density units ($\Sigma_{crit}$), where
$\Sigma_{crit} = \frac{c^2}{4\pi G}\frac{D_{s}}{D_{l}D_{ls}}$
and $D_{s}, D_{l}$ and $D_{ls}$ correspond to the angular diameter distance between the observer and the source, the observer and the lens (the cluster) and the lens and source, respectively. CLASH convergence maps have been
scaled to $D_{s}/D_{ls} =1$ \cite{zitrin1}, therefore, the $\kappa$-map for each cluster can be easily converted into a $\Sigma$-map. 

In order to reconstruct the three dimensional density profiles of the clusters, we assume spherical symmetry and fit the analytical projected Navarro-Frenk-White (NFW, \cite{nfw,bartelman}) density profile to the radially averaged profiles of the $\Sigma$-maps. We perform a Markov Chain Monte Carlo (MCMC) fit to get the best-fit values of the free parameters $r_s$ and $c_{200}$, the characteristic cluster radius and the concentration, respectively. Finally, we can reconstruct the lensing mass profiles by spherically integrating the best-fit NFW density profile for each cluster. For some clusters in our sample, two different models of $\kappa$-maps were reconstructed, Light-Traces-Mass and Pseudo-Isothermal Elliptical Mass Distribution with elliptical NFW, hereafter LTM and PIEMD+eNFW. Details can be found in \cite{zitrin1,zitrin2, jullo,nfw}.
For a coherent comparison with the HSE mass profiles, surface mass density profiles are centered in the same positions as the pressure and electron density profiles in Section~\ref{hse}. As a consequence, the centers of some projected density profiles may not coincide with the density peaks in the maps.


\section{Hydrostatic and lensing mass comparison}
\begin{figure}[h]
    \centering
    \includegraphics[trim={10cm 0cm 10cm 3cm},scale=0.25]{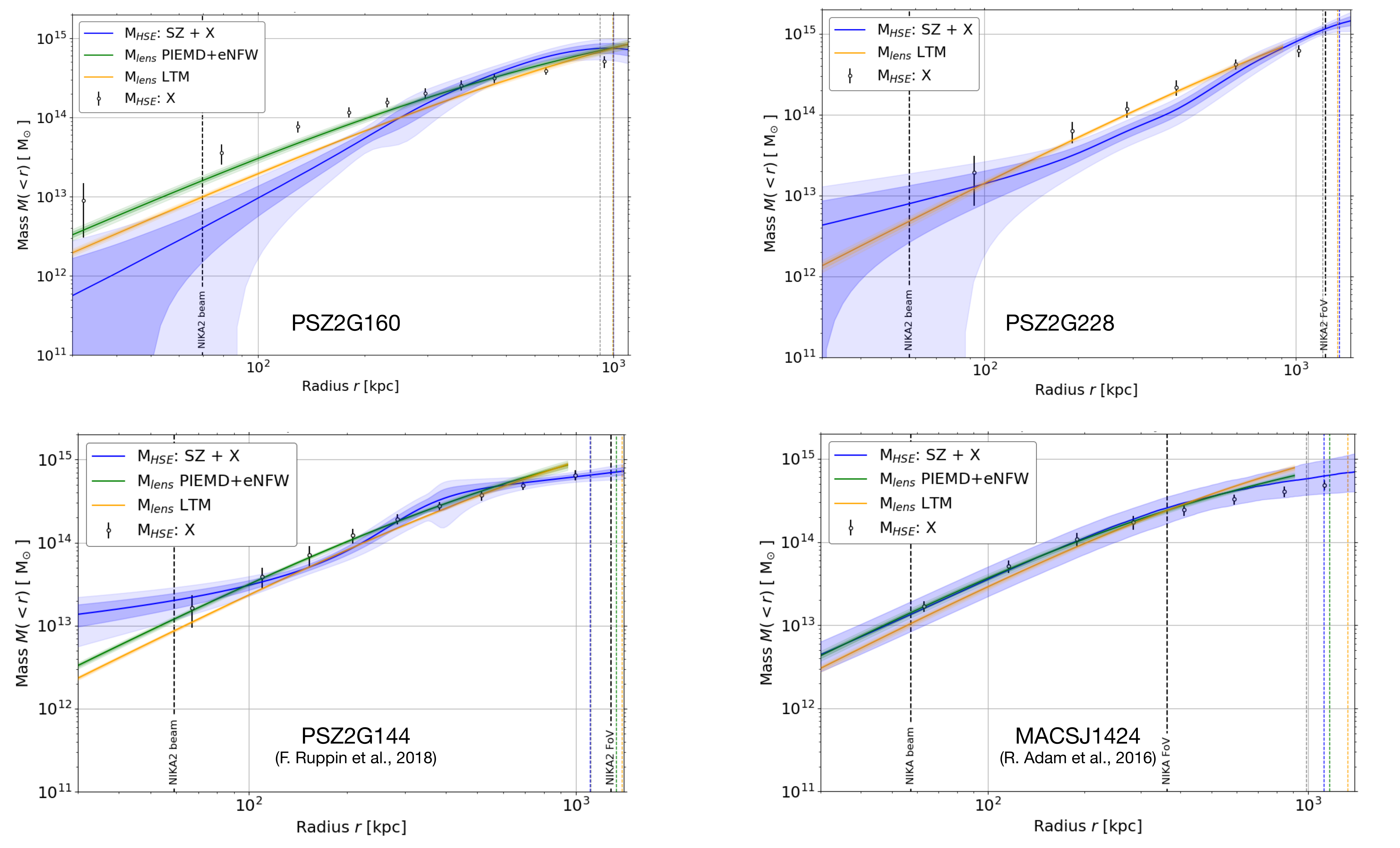}
    \caption{Reconstructed HSE mass profiles: in blue for SZ+X-ray combination and in white dots for X-ray only. Reconstructed lensing mass profiles for the two $\kappa$-map models in green and yellow. Vertical colored dashed lines represent the $R_{500}$ for each mass profile and black dashed lines NIKA(2) instrumental beam and field of view.}
    \label{massprofiles}
\end{figure}

The reconstructed HSE and lensing mass profiles for PSZ2-G160.83+81.66, PSZ2-G228.16+75.20, PSZ2-G144.83+25.11 and MACS J1423.8+2404 are presented in Figure~\ref{massprofiles}, with $1\sigma$ and $2\sigma$ confidence envelopes. The envelopes of the lensing mass profiles are very thin compared to the HSE ones. This is due to the small amount of parameters (2 in the NFW, instead of more in \cite{umetsu2014,umetsu2018}) that we fit compared to the case of the HSE mass, as well as the fact that we consider the convergence maps as true, meaning that the radially averaged profiles of the $\Sigma$-maps that we use for the fit only contain statistical uncertainties. 
Nevertheless, the uncertainties in our results are consistent with the ones in \cite{pennalima} and we will proceed with them. Due to the complex morphology of MACS J0717.5+3745 and the simplicity of our model, we choose to analyse this cluster with a different approach in Section 5.


We concentrate here in comparing the different mass estimates obtained for the cluster sample at the characteristic $R_{500}$ radius. In Figure~\ref{mass500} we show the probability distribution in the plane $M_{500}^{\rm{HSE}}$ and $M_{500}^{\rm{lens}}$ for the four clusters. The $M_{500}^{\rm{HSE}}$ are obtained from SZ and X-ray. The lensing masses are derived from the combination of the two lensing mass estimates (except for PSZ2-G228.16+75.20 for which we only have one lensing mass), to account for modeling effects. For PSZ2-G160.83+81.66 both $M_{500}^{\rm{lens}}$ are in agreement and consistent with the $M_{500}^{\rm{HSE}}$. The complex morphology of PSZ2-G228.16+75.20 \cite{bonafede2} makes the spherical modeling results difficult to interpret. Even if our $M_{500}^{\rm{HSE}}$ estimate is consistent with \cite{psz2cat}, a more thorough study of this cluster may be needed. For PSZ2-G144.83+25.11 the SZ data (see \cite{ruppin1}) is of very good quality and the spherical electronic pressure profile, having masked the overpressure region, fits very nicely the data. This gives a very well defined HSE mass estimate. The two $M_{500}^{\rm{lens}}$ estimates for PSZ2-G144.83+25.11  are in agreement within $1\sigma$ uncertainties, but the differences between the $\kappa$-map models enlarges the error bar in the final $M_{500}^{\rm{lens}}$ distribution. An extreme case of this effect is seen for MACS J1423.8+2404: the $M_{500}^{\rm{lens}}$ estimation for each of the lensing models is so different that the combined distribution reveals two distinct peaks. These results highlight the impact that systematics can have in our mass estimate. We see that the data quality, the cluster morphology and dynamical state, as well as the chosen observables and the modeling influence the mass estimate.
\subsection{Hydrostatic-to-lensing mass bias}
\begin{figure}[h]
    \centering
    \begin{minipage}[b]{0.6\textwidth}
        \includegraphics[scale=0.2]{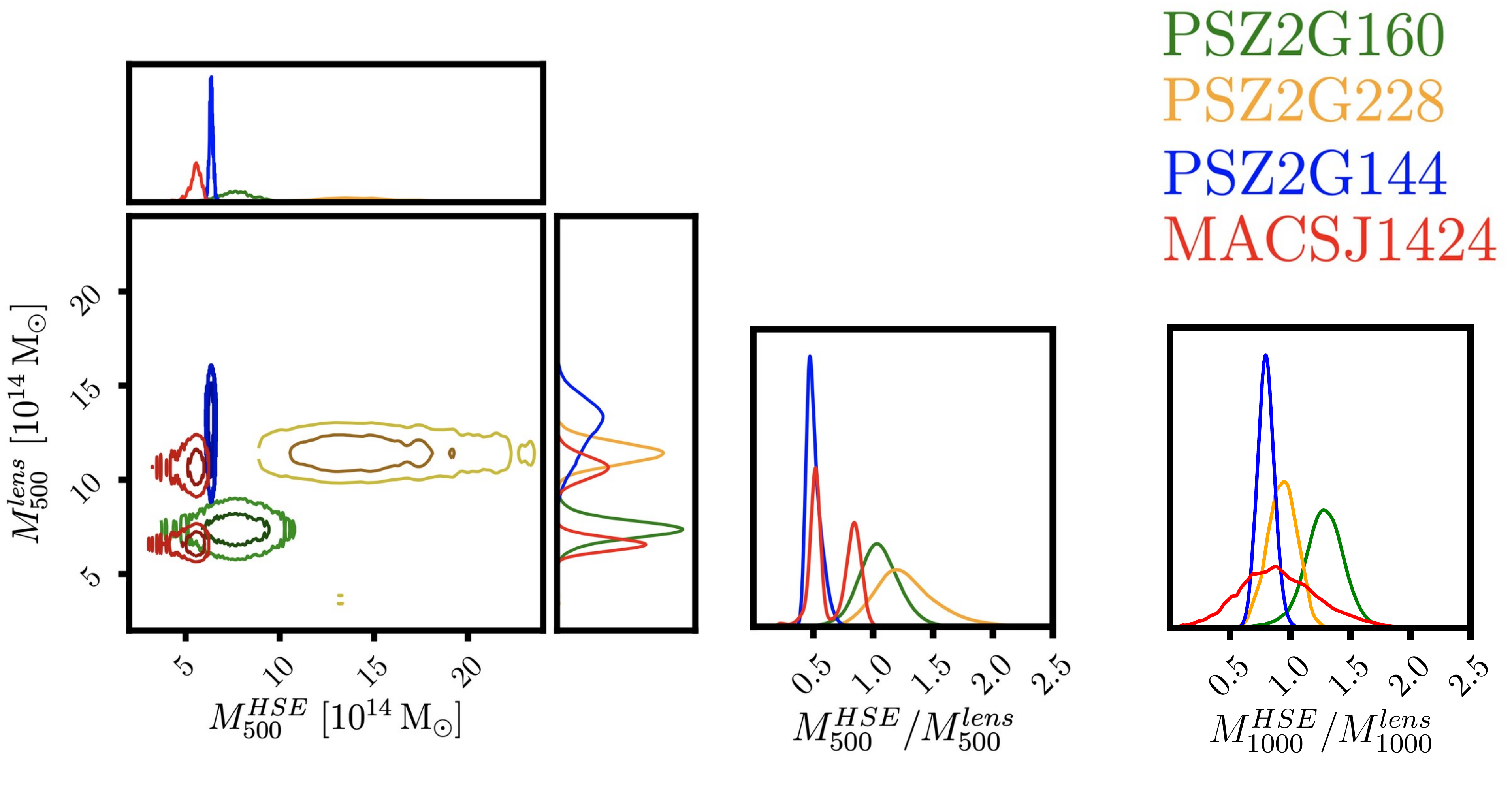}
        \caption{\textit{Left:} marginalized distributions of the density probability for the lensing and HSE mass at $R_{500}$. \textit{Center:} hydrostatic-to-lensing mass ratio for each cluster at $R_{500}$. \textit{Right:} hydrostatic-to-lensing mass ratio for each cluster at a fixed $R_{1000}^{\rm{HSE}}$.} 
        \label{mass500}
    \end{minipage}
    \hfill
    \begin{minipage}[b]{0.35\textwidth}
        \includegraphics[trim={0cm 0cm 0cm 0cm}, scale=0.1]{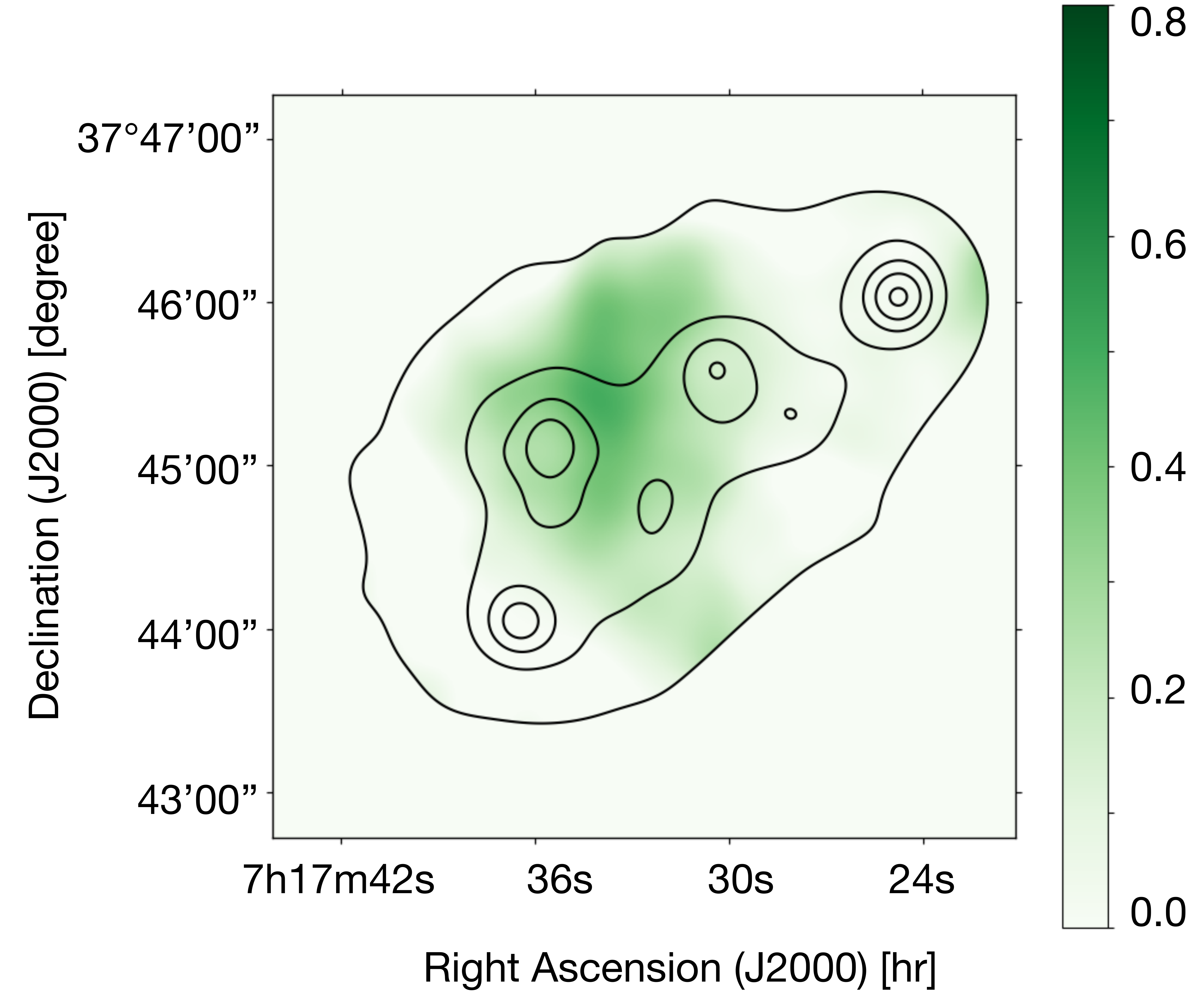}
        \caption{tSZ/$\kappa$ map for MACS J0717.5+3745. Map pixels for which $\kappa$ < 0.5 are set to zero. Contours correspond to [0.5, 1, 1.5, 2, 2.5] levels in $\kappa$-map. The map has been smoothed with a 18’’ Gaussian kernel.}
        \label{ratiomap}
    \end{minipage}
\end{figure}

The hydrostatic-to-lensing mass bias, $b_{\rm{HSE}/\rm{lens}} = 1 - M_{500}^{\rm{HSE}}/M_{500}^{\rm{lens}}$, is computed from the distributions in the left panel in Figure~\ref{mass500} and presented in the center as the ratio between the HSE and the lensing mass. The results vary largely from cluster to cluster and systematic effects in mass estimates propagate to the bias. The mean hydrostatic-to-lensing mass ratio of this small sample is $M_{500}^{\rm{HSE}}/M_{500}^{\rm{lens}} = 0.87 \pm  0.47$. 
This is in agreement with hydrodynamic simulations \cite{gianfagna}, but we notice that the dispersion in our result is very large. This is also consistent with the value needed to reconcile the cosmological constraints obtained from Planck CMB power spectra to the cluster counts,  $M_{500}^{\rm{HSE}}/M_{500}^{\rm{tot}} = 0.58 \pm 0.04$ \cite{planck14, planck1}. For such comparisons we have to consider the lensing mass as an accurate estimate of the total mass of clusters.

We have also compared the hydrostatic-to-lensing mass bias in a common inner region ($R_{1000}$) to avoid any extrapolation of the lensing mass profile. We show in the right panel in Fig. \ref{mass500} the ratios at the $R_{1000}$ fixed by the SZ+X-ray HSE mass profile. In this case from the combination of the four clusters, we obtain a mean value of $M^{\rm{HSE}}/M^{\rm{lens}} (R_{1000}^{\rm{HSE}}) = 0.97 \pm  0.29$. Notice that this ratio can not be directly compared to the $R_{500}$ one as we expect some variations with radius.

\section{Identification of structures by comparing SZ and lensing maps}
When comparing HSE and lensing mass results with spherical profiles we mix up different effects that can hardly be disentangled and may bias the final results. This motivates a map to map comparison for complex clusters. We take as a test case the MACS J0717.5+3745 galaxy cluster. Previous studies \cite{adam3,adam4,adam5,bonafede} have revealed its complex morphology and dynamical state: at least 4 sub-clusters interacting in a complex merger scenario, as well as sub-clusters moving along the line-of-sight direction. We combine here the tSZ map from \cite{adam3} and the $\kappa$-map from CLASH \cite{zitrin1}. In Figure~\ref{ratiomap} we present the tSZ/$\kappa$ ratio map (in an arbitrary color scale), with contours showing the levels in the $\kappa$-map. The tSZ map has been normalized to $Y_{500}$ and the $\kappa$-map to $M_{500}$. The darkest green regions indicate the areas where the relative contribution of the thermal pressure from the hot gas to the total mass budget is greater than average. The presence of such hot region is consistent with outcomes from X-rays and from the combination of X-rays and tSZ data \cite{adam5}. 


\section{Conclusions}

In this work we have constructed a NIKA and NIKA2 LPSZ-CLASH common sample of five high redshift (0.55 - 0.89) galaxy clusters. We have compared for the first time their HSE and lensing mass estimates, showing the impact that observables have in the mass calculation. Moreover, we have highlighted the important contribution of lensing models to the mass uncertainties and we have accounted for them. From the HSE and lensing mass profiles reconstruction, we have obtained the hydrostatic-to-lensing mass bias for each cluster in the sample. We observe a large dispersion across the sample and hints of contribution from systematics. It will be, therefore, essential to treat these effects carefully when dealing with larger samples. From this sample, we can not draw conclusions about the redshift evolution of the bias.
In addition, the analysis performed on MACS J0717.5+3745 has illustrated the potential of map to map tSZ and lensing comparison for revealing cluster physics.

\section{Acknowledgements}
\scriptsize{We would like to thank the IRAM staff for their support during the campaigns. The NIKA2 dilution cryostat has been designed and built at the Institut N\'eel. In particular, we acknowledge the crucial contribution of the Cryogenics Group, and in particular Gregory Garde, Henri Rodenas, Jean Paul Leggeri, Philippe Camus. This work has been partially funded by the Foundation Nanoscience Grenoble and the LabEx FOCUS ANR-11-LABX-0013. This work is supported by the French National Research Agency under the contracts "MKIDS", "NIKA" and ANR-15-CE31-0017 and in the framework of the "Investissements d’avenir” program (ANR-15-IDEX-02). This work has benefited from the support of the European Research Council Advanced Grant ORISTARS under the European Union's Seventh Framework Programme (Grant Agreement no. 291294). F.R. acknowledges financial supports provided by NASA through SAO Award Number SV2-82023 issued by the Chandra X-Ray Observatory Center, which is operated by the Smithsonian Astrophysical Observatory for and on behalf of NASA under contract NAS8-03060. }



%
%
%

\end{document}